\documentstyle[prl,aps,psfig,twocolumn]{revtex} 
\newcommand{\binom}[2]{\left( \begin{array}{clcr} {#1}  \\
                                                  {#2}  \end{array} \right) }
\newcommand{\avg}[1]{\langle {#1} \rangle}
\begin{document}
\twocolumn[\hsize\textwidth\columnwidth\hsize\csname@twocolumnfalse\endcsname
\title{Growing Dynamics of Internet Providers.}
\author{Andrea Capocci$^1$, Guido Caldarelli$^2$, Riccardo
Marchetti$^2$ and
Luciano Pietronero$^2$.}
\address{$^1$Institut de Physique Th\'eorique,
Universit\'e de Fribourg, CH-1700, Fribourg, Switzerland.}
\address{$^2$INFM Sezione di ROMA1 Dip. Fisica, Universit\`a di
Roma
``La Sapienza'' P.le A. Moro 2 00185 Roma, Italy.}
\date{\today}
\maketitle

\begin{abstract}
In this paper we present a model for the growth and evolution of
Internet providers.
The model reproduces the data observed for the Internet connection
as probed by tracing routes from different computers.
This problem represents a paramount case of study for
growth processes in general, but can also help in the 
understanding the properties of the Net. 
Our main result is that this network can be reproduced by a
self-organized interaction  between users and providers that
can rearrange in time.
This model can then be considered as a prototype model for the
class of phenomena of aggregation process in social networks.
\end{abstract}
\pacs{05.40-a, 87.23.Ge, 02.10, 05.70.Jk}
]
\narrowtext

Networks are systems composed by elementary units, the {\em
nodes}, connected by  directed or undirected {\em links}.
The number of links pointing to a node, $k$, is known as the
degree of the node, whose distribution gives the network {\em connectivity}.
This simple structure is almost ubiquituous in Nature, and
the reason of such a success is often linked to the optimization
of some cost function. For example, in all transport processes
networks are selected to efficiently distribute the quantities of
interest among the sites connected.
Networks could also be used to describe both the spreading of
information or diseases \cite{Wat2} and physical structures as, 
for example, river basins \cite{rin}, biological distribution 
networks (vascular systems)\cite{ban} and some properties of the 
hardware layout of Internet\cite{fal3,Gca}. A detailed discussion of
such networks and some models are described in Ref.\cite{Sca}. 

Here, we present some experimental measures of the network of
Internet providers and we propose a simplified model in order to
explain them.
It is worth to note that this network does not correspond to the 
one composed by the web pages. This network is composed by
the physical connections of the computers and the 
measures come from the analysis of the data provided by
the Internet Mapping Project \cite{IMP} Hereafter we are going
to discuss only this particular system and we do not want to describe
neither web pages network nor other social systems.
Recently\cite{Gca}, some statistical
properties  of the connectivity of this physical network of
Internet have been investigated. For such a system, a tree-like
structure has been found by checking the routers connections
from a starting point. Despite the bias
introduced by observing the Net from a single node, some
statistical feature can be established, as the power law
distribution of the degree.
Here, instead we are focusing on the possible dynamics behind
the formation of such a structure.
The main results from the data analysis is the power-law
distribution of site degree showing the absence of a particular scale.
It would be tempting then, to assume that such scale-free
distribution has been
originated by some sort of optimization of the supply present in
the providers market. This is the main idea inspiring our dynamical
model that should mimic the evolution of a system of users and
providers.
The model we propose here is in close relationship with a
prototype growth model introduced by Simon \cite{Sim} and recently 
improved \cite{Sca,Bar} in order to
explain the widespread occurrence of fractal behaviour in several cases
ranging from the web-pages statistics \cite{Bor,Dor} to scientific
citation\cite{new} actors in the same movie cast \cite{Hub,Bar2,Bar3,Wat}.
Some of the networks considered displays these scale-free
properties, as a result of some optimization, as, for example, for the blood
vessels \cite{ban} or the river basins \cite{mar}.
In others, a "Small World" phenomenon arises \cite{Wat2}, and through
suitable shortcuts all the points are connected one each other in few steps.
Together with the numerical analysis upon real social networks,
a strong effort
is provided by the physicists' community to find suitable
theoretical models for such systems.

\begin{figure}
\centerline{\psfig{file=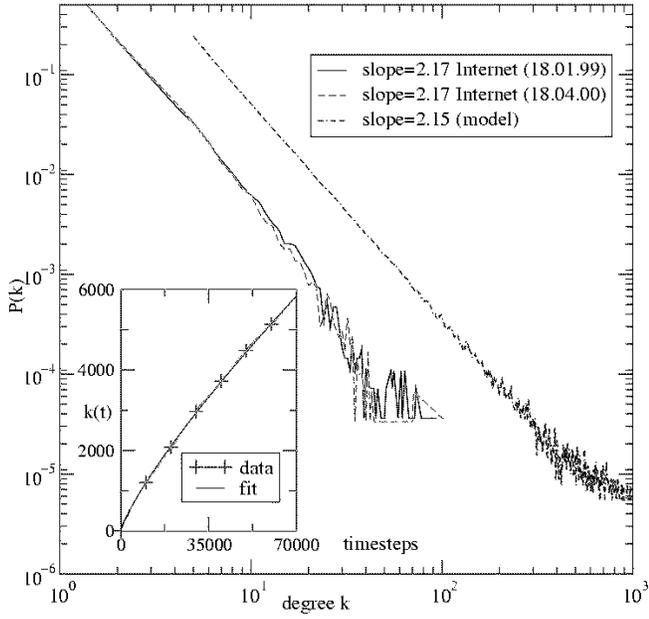,width=9cm,angle=0}}
\caption{The degree distribution in the model with $r=0.29$ and $d=0.41$ and
in the real Internet data collected on different days. In the inset there is 
the temporal behavior of the degree of a provider.} 
\label{fig1}
\end{figure}

The set of data is obtained through a computer
instruction that allows to trace the route from one terminal to
any allowed
address in the Internet domain. The $UNIX$ command {\em
traceroute} records
all the nodes through which the target is reached from the
starting point
where the command is run.
These paths can change over time for the following reasons.
Firstly the routes reconfigure since the path is variable
according to the traffic at the moment or more generally according to the
availability
of the connection.
Secondly the whole structure is physically evolving due to the
new connections
that take place.
Nevertheless the main statistical properties of this structure
remain constant in time even if the total number of connections
increases.
These data can be put in a tree-like structure such that
providers are organized in levels: the main providers on
the top level are linked to secondary providers, that provide the
connection to
successive levels down to the common user level.
The degree of the providers can now be computed over all
the levels of the network.
The main result is that the Probability Density Function
(PDF) to find a node with degree $k$ scales following a power
law (see Fig.\ref{fig1})
where the exponent $\gamma$ is equal to $2.2$.
\begin{equation}
P(k)=k^{-\gamma}
\end{equation}
Since a similar value, is also known to
describe the power-law distribution of links in web pages,
it is possible that a similar evolution holds for both of them.

In particular, we propose a mechanism that describes the
development of
the connections between two subsequent levels in a network.
In our model, two different classes of nodes are present,
representing providers and users (that, possibly, could act
as providers for a lower level of users).  
Sites representing providers can have several links, pointing 
to other sites corresponding to users.
Users, on the other hand, have a single link, pointing
to their provider. They are not allowed to have more than one
provider.
By iterating this microscopical interaction level by level one
could, in principle, recover the whole tree-structure of the network.
At each time step, a node is added to the network.
The new node can be either a provider with a probability $r$ or
a user with probability $1-r$.
When a provider is added, $D(t)$ users in the network are chosen
at random, and rewired to the new provider.
Links to the previous providers are then removed.
We assume that the integer number $D(t)$ is a random variable 
with Poisson distribution and mean value $d$.
This aims to mimic the fact that a real provider decides to 
enter the network when it expects to acquire a certain number of 
connected users, $d$ on average, according to some microeconomical 
optimization rule. 
The randomness of $D(t)$ takes into account inexact forecasts about 
the number of rewiring users.

This addition of a provider does not change the total number of
links in the network. Instead, when a user is added
it is linked through a new link to an existing provider.
Then, the addition of a user increases by one the total number
of links.
The probability that a provider acquires a new user is
proportional to
its degree, that is, the number of users it is linked to.
This rule known as "richer gets richer condition" is at the
basis of the typical behaviour observed in scale-free 
networks\cite{Bar,Bar2}
differently from the features shown by ordinary random
graph\cite{ER}.
We call $k_{i}(t)$ the degree of the $i$-th provider (introduced 
at time $t_{i}$) and
\begin{equation}
K(t) = \sum_{i}k_{i}(t)
\end{equation}
the total number of links (and users) in the network, at time $t$.
A user is added at a rate $(1-r)$ per time step and is connected
to a provider with probability proportional to its degree.
Then, the $i$-th provider acquires a new link with a probability
$(1-r)k_{i}/K$.
A provider is added with probability $r$ at each time step. Each
user has the same probability $1/K$ to be rewired to the new provider.
Thus, a provider with degree $k_{i}$ loses $l$ users ($l>0$) with
``binomial'' probability $P_C(l)$
\begin{equation} \label{bind}
P_C(l)=r\binom{D(t)}{l} (k_{i}/K)^{l}(1 - k_{i}/K)^{D(t)-l},
\end{equation}
whose mean value is $D(t)k_{i}/K$.
The degree of a provider does not change with the
remaining
probability $(1-r) (1-k_{i}/K)+r(1 - k_{i}/K)^{D(t)}$.
Since new links are created at rate $(1-r)$ per time step, the
number of
links at time $t$ is $K(t)=(1-r)t$, for large $t$ values.
Thus, one can compute the time evolution of the average
connectivity $\overline
{k_{i}}(t)$ over many realizations of the model.
To do that, we assume that the correlation between $k(t)$ and
$D(t)$ can
be neglected and that the two average can be taken
independently and $D(t)$ be replaced by $d$ in the mean value of eq. (\ref{bind}):
\begin{equation} \label{motion}
\overline{k_{i}}(t+1) =
\overline{k_{i}}(t) + (1-r)\frac{\overline{k_{i}}(t)}{K(t)}
-rd\frac{\overline{k_{i}}(t)}{K(t)},
\end{equation}
where the second term in the right hand side of this equation
corresponds to the addition of a new user, and the third term 
corresponds to the subtraction of links after the birth of a 
new provider. 
This equation can be written in the continuous limit as
\begin{equation}
\frac{d\overline{k_{i}}}{dt} =
\frac{1-(d+1)r}{(1-r)t}\overline{k_{i}};
\end{equation}
This, with the boundary condition $k_{i}(t_{i})=d$, gives
\begin{equation} \label{tscale}
\overline{k_{i}}(t) = d\left(\frac{t}{t_{i}}\right)^{\sigma}
\end{equation}
where
\begin{equation}
\sigma = \frac{1-(d+1)r}{1-r}.
\end{equation}
One can see that $k_{i}(t)/K(t)$ goes to $0$ as $t$ goes to
$\infty$ for all $i$, showing that no node grows in degree 
as fast as the whole network. 
The stability of the network is then assured.
The dynamical behavior described in equation (\ref{tscale}) is in good 
agreement with the numerical simulations of the model,
whose dynamical properties are shown in the inset of Fig.\ref{fig1} for a
single provider. 
By means of this relation between time and degree, we can
now compute the probability that a provider has a degree 
less than $k$,
$P(k_{i} < k)$.
We assume that $P(k_{i} < k) \simeq P(\overline{k_{i}} < k)$.
Solving equation (\ref{tscale}) for $t_{i}$, one can see that
\begin{equation}
\overline{k_{i}(t)} < k
\Longleftrightarrow \frac{t_{i}}{t} >
\left(\frac{k}{d}\right)^{-\frac{1-r}{1-(d+1)r}}.
\end{equation}
This means that providers with a degree less than a given
value $k$ are those ones which have been added to the
network after a corresponding time, and have not had time enough
to develop a cluster of $k$ users around them. Since
nodes are added at a uniform rate,
\begin{eqnarray}
P(k_{i} < k) & \simeq & P ( t_{i}/t > (k/d)^{-\frac{1-r}
{1-(d+1)r}} ) = \\ \nonumber
& = & 1 - \left(\frac{k}{d} \right)^{-\frac{1-r}{1-(d+1)r}}.
\end{eqnarray}
We can then write for the PDF $p(k) = dP(k_{i}<k)/dk$, which
yields
\begin{equation}
p(k) \sim k^{-\gamma};
\end{equation}
where
\begin{equation} \label{gamma}
\gamma = \frac{2-(d+2)r}{1-(d+1)r}.
\end{equation}
\ref{gamma} provides us with an upper bound on $r$, 
since one can see that the exponent diverges at $r = \frac{1}{d+1}$.
Numerical simulations, as shown in Fig.\ref{fig1}, follow 
the predicted behavior.
We recall here that the external parameters $r$ and $d$ are
estimated by statistical surveys of the Internet.
Through the {\em traceroute} procedure one can describe the
connection to the outlet by means of a tree-like structure.
However, the iteration of this procedure does not show the whole
structure of a given region of Internet, since cross-links
between sites at the same distance are not seen.
Yet, some statistical property of the considered network can
still be established.
We assume that {\em traceroute} shows only a given fraction 
$\mu < 1$ of the real number $k$ of links pointing to a site.
This, however, does not affect the shape of the distribution 
(if it is a scale-free one) and the reliability of our 
statistical survey. 
We then call $k_{eff} = \mu k$ the apparent degree of a site.
By the {\em traceroute} picture of the physical network
of Internet, the degree
distribution density $q(k_{eff})$ shows a power-law behavior
\begin{equation}
q(k_{eff}) \sim k_{eff}^{-\alpha}
\end{equation}
where $\alpha \simeq 2.2$.
For the considerations made above, the exponent found by the {\em
traceroute} analysis is a good approximation of the real value.
This value is slightly different from the $2.48$ recovered by the
analysis of Ref.\cite{fal3}. We believe that this difference arises 
mainly from the growth of the Internet, (that is now very different from that at 
the time).
This enables us to write $\gamma \simeq 2.2$.
The connectivity $\avg{k}$ cannot be computed by {\em
traceroute}, since the fraction $\mu$ is unknown.
Nevertheless, $\avg{k}$ is provided in other published
statistical analysis \cite{Nla}, according to which the 
connectivity is $3.4$.
Nevertheless, we checked that for the first
layers ifo the data analyzed the measured value is not that far
from the above one. We also notice a decrease of the 
connectivity with the distance from the source of traceroute.
We decide here to focus on the first levels that can be effectively 
probed by this analysis.
If we assume that our model describes the way a network is built
at each level, the predicted value for $\avg{k}$ is
\begin{equation}
\avg{k} = 1 + \frac{1-r}{r}.
\end{equation}
The unity in the right hand side takes into account that each
site has a provider and a link that points to it, this is not considered
in our model and must be explicitly added.
The second term is the ratio between users and providers in our
model.

This equation provides us with the value of $r \simeq 0.29$.
Replacing this value into equation (\ref{gamma})
one can recover the value of $d \simeq 0.41$.
Such a value of $d$, smaller than $1$, shows that our model
describes the real structure of Internet when some provider is
introduced without rewiring any user, as it is suggested by 
the third term in the right
hand side of equation (\ref{motion}). 
If a new providers is born and no user get rewired, the provider 
is sentenced to death, since a provider without users cannot survive.

Until now computation has been done in the limit hypothesis of connection
between users and only one provider. 
One can study through numerical simulation the behaviour when users are 
allowed to be linked with different users. 
In this case, when a provider is added, users rewired to it
keep their old provider connection.

\begin{figure}
\centerline{\psfig{file=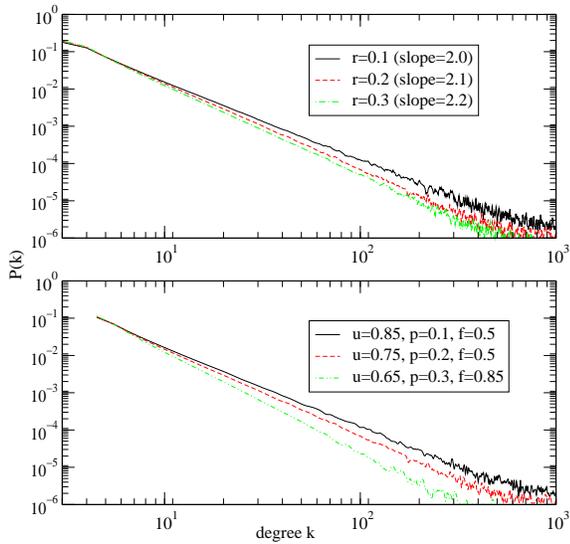,width=8cm,angle=0}}
\caption{(above) Degree distribution for different values of the providers birth rate 
$r$ allowing users to have more than one provider.
(below) Degree distribution with different user birth rates $u=1-r$ and 
probability of merging $f$.}
\label{fig3}
\end{figure}

In our model, the possibility to be connected to several users corresponds to
neglecting the third term in equation (\ref{motion}), which takes 
into account the probability for a provider to lose a user due to a 
newborn provider, and replacing $K(t)$, the total number of links, by 
\begin{equation}
K(t)=[1+r(d-1)]t,
\end{equation}
since now one link is added when a user is added, and $d$ links, on average, 
when a provider is added.
Performing the same computation as above, one would expect to obtain a 
scale-free degree distribution, with an exponent 
\begin{equation}
\gamma = \frac{2+r(d-2)}{1-r}.
\end{equation}
This behavior is confirmed by simulations, as can be seen in Fig.\ref{fig3}.
In addition, we simulated the case in which providers merge at a uniform 
rate. 
We assume that at each time step, providers are added at a rate 
$r$, with a probability $f<r$ a randomly chosen provider 
vanishes and users connected to it are rewired to another provider, 
according to the ``richer-gets-richer'' rule, and users are 
introduced with probability $u=1-r-f$.
The assumption made on $f$ is needed to avoid the extinction of almost 
all provider as each merging decreases by $1$ the number of provider.
If the merging rate is higher than the birth rate of new providers
the number of providers rapidly tends to $1$.
As well as in the previous versions of the model, this growing network 
displays a scale-free distribution of degree. 
This result is shown in the lower part of Fig.\ref{fig3}, where we plotted the degree 
distribution for different values of $r$ and constant $f$.
This scale free behaviour characteristic of ``social'' networks,
has been recently explained\cite{Bar2} by means of two ingredients:
firstly, the number of nodes has to grow in time and secondly the
nodes with greater degree are advantaged in acquiring
new links. 
This model gives  an exponent $\gamma = 3$, while
the real exponents found in the social
networks considered above are in the range between $2$ and $3$.
Since then, other models of growing networks have been proposed, whose
degree distribution are closer to the real ones in the
corresponding real network \cite{Dor2,Kra}.
The model we introduced describes the dynamical development of a
network composed by two classes of nodes, as it is the case in the
Internet connections between providers and users.
In fact, the physical structure of Internet is made of
superposed levels of nodes, corresponding to providers,
subproviders, or users at the lowest
level, whose distribution of degree has been recently
found to show a power law behavior.
The model exhibits the same scale-free shape depending on the
external parameters $r$, i.e. the providers fraction in the
total number of nodes, and $d$, the average number of users
who join a new born provider.
The parameters can be naturally tuned to realistic values
to recover the exact exponent of the tail of the
distribution of the degree.

We acknowledge the support of EU contract FMRXCT980183.

\newpage

\end{document}